\documentclass[preprints,article,accept,moreauthors,10pt,a4paper]{mdpi} 

\firstpage{1} 
\pubvolume{xx}
\issuenum{1}\articlenumber{1}
\pubyear{2018}
\copyrightyear{2018}
\externaleditor{Academic Editor: name}
\history{Received: date; Accepted: date; Published: date}

\Title{Workshop summary ``The Power of Faraday Tomography''}

\Author{Marijke Haverkorn $^{1,}$*\orcidA{}, Mami
  Machida $^{2}$ and Takuya Akahori $^{3}$}

\AuthorNames{Marijke Haverkorn, Mami Machida and Takuya Akahori}

\address{%
$^{1}$ \quad Department of Astrophysics / IMAPP, Radboud University Ni
jmegen, PO Box 9010, 6500 GL Nijmegen, The Netherlands; m.haverkorn@astro.ru.nl\\
$^{2}$ \quad Department of Physics, Faculty of Sciences, Kyushu
University, 744 Motooka, Nishi-ku, Fukuoka, 819-0395, Japan;
mami@phys.kyushu-u.ac.jp\\
$^{3}$ \quad Mizusawa VLBI Observatory, National Astronomical
Observatory of Japan, 2-21-1 Osawa, Mitaka, Tokyo 181-8588, Japan}

\corres{Correspondence: m.haverkorn@astro.ru.nl}

\abstract{This article summarizes the work presented at the workshop
  ``The Power of Faraday Tomography: towards 3D mapping of cosmic
  magnetic fields'', held in Miyazaki, Japan, in Spring 2018. We place
  the various oral and poster presentations given at the workshop in a
  broader perspective and present some highlight results from every
  presenter.}

\keyword{cosmic magnetism; Faraday Tomography; galactic magnetic
  fields; MHD}

\begin{document}

\section{Introduction}

Magnetic fields play a vital role on all scales throughout the
Universe: allowing the creation of stars and exoplanets, affecting the
gas flows in the interstellar medium, forming galactic and AGN jet
structures, accelerating cosmic rays, and permeating the cosmic web
and galaxy clusters. Yet the origin of these cosmic magnets and the
mechanisms of field amplification and ordering over the history of the
Universe are still largely unsolved.

“Cosmic Magnetism” is recognized as one of the key science topics for
the largest radio facilities such as the Low-Frequency Array (LOFAR),
the Karl G. Jansky Very Large Array (JVLA), and Atacama Large
Milimeter/submilimeter Array (ALMA), as well as the Square Kilometre
Array (SKA) and its precursors Murchison Widefield Array (MWA),
Hydrogen Epoch of Reionization Array (HERA), the Australian SKA
Pathfinder (ASKAP), and MeerKAT. We are now entering the era of these
sensitive and high-resolution facilities, which are expected to
uniquely solve many outstanding questions in cosmic
magnetism. Theoretical and numerical predictions will become much more
important in this era.

One of the technological breakthroughs in modern radio telescopes is
wide bandwidth in frequency. It improves, for example, the
sensitivity, the spectral index estimation, and depolarization
analysis. Moreover, it brings us an innovative data analysis method
called Faraday Tomography. Here, wide frequency coverage means a “big
data” challenge: computational cost is a major issue which should be
addressed now to resolve it by the time when the future largest
projects run.

The Japanese workshop organizers, part of the Japan SKA Consortium
(SKA-JP) Cosmic Magnetism Science Working Group, actively address the
(future) capabilities of Faraday tomography \cite{akahori2016,
  akahori2018}. The organizers recently had an opportunity to start a
collaboration between the SKA-JP and a group in the Netherlands, and
put minds from both groups to the problems described above.

This
collaboration resulted in the workshop ``The Power of Faraday
Tomography: towards 3D mapping of cosmic magnetic fields'',  held in
Miyazaki, Japan, from 28 May to 2 June 2018. The goal of the workshop
was to present and discuss new results on cosmic magnetism research,
with an emphasis on the analysis method of Faraday Tomography. The
workshop was meant to be interactive and partly educational:
presentations ranged from instructive reviews by senior scientists to
new results to tutorials on various analysis methods \cite{ideguchietalFT}. A meeting of the
POlarisation Sky Survey of the Universe's Magnetism (POSSUM) meeting
was held at the workshop, which is not included in these proceedings.

The workshop had 64 participants. The biggest group of these were
Japanese (24), there were 19 from Europe (6 Germany, 4 Netherlands, 4
Italy, 2 UK, 1 France, 1 Spain), 8 from North America (5 Canada, 3
US), 8 from Australia, and 4 from Asia other than Japan (2
South-Korea, 1 Russia, 1 India). A relatively large part of these were
young researchers: 28 postdocs, 12 PhD students and 2 MSc students.

We emphasize that this article does {\it not} contain a full
discussion of the scientific fields discussed, but only describes the
contributions to the various fields made in this workshop. In
addition, we do not try to summarize all scientific results in every
contribution, but give a few highlights and try to embed the topic
into the bigger picture. We encourage the reader to turn to the full
proceedings article of any contribution that catches your interest in
this summary.
 
%

\section{Summary of the conference}

The workshop contributions were ordered in seven sessions on related
scientific topics. In this section, we briefly introduce the topic and
discuss the presentations given in each of these sessions.


\subsection{Faraday Tomography}

Faraday tomography is a relatively new tool to study magnetic fields
in galaxies, which allows decomposing various synchrotron emission
components along a line of sight into their various Faraday rotation
contributions. At the conference,
{\bf Shinsuke Ideguchi} presented simulated Faraday spectra of face-on
nearby galaxies with varying regular magnetic field strengths and
varying turbulent field coherence lengths \citep{ideguchietal2017}. He
showed that the width of the Gaussian Faraday spectrum of the
turbulent fields varies with coherence length of the field, and that
its shape is altered with varying regular fields. Therefore, measuring
the moments of Faraday components in a spectrum gives information on
the magnetic field in a spiral galaxy.

Broadband data is extremely important for Faraday Tomography both for
Faraday depth resolution as well as sensitivity to Faraday thickness;
{\bf Yoshimitsu Miyashita} shows that while in a limited radio band
the data and a model can agree, they can diverge widely outside a
narrow observed band \cite{miyashiteetal2016}, giving rise to false
trust in well-fitting but wrong results.

\subsubsection{Alternative methods for radio-polarimetric data
  analysis}

{\bf Shinsuke Ideguchi} reviewed and explained the method of {\it
  QU-fitting} \cite{ideguchietal2014, ideguchiFT}, in which the observed
wavelength dependent behavior of Stokes Q and U is compared to various
models of components of the Faraday-rotating magnetized gas.

{\bf Dmitry Sokoloff} combines Faraday Tomography with wavelet
analysis to study Faraday depth fluctuations at varying resolutions,
in order to render coherent fields in spiral arms visible in Faraday
depth maps with dominating small scale Faraday depth variation
\cite{sokoloffetalFT}.  With these combined methods, it is possible to
identify typical magnetic arms (in the stellar interarm regions) as
observed in some nearby spiral galaxies.

\subsubsection{Other advances in radio (spectro-polarimetry)
  calibration techniques}

{\bf Wasim Raja} emphasized the importance
of polarization calibration, especially off-axis. He introduced the
Australian Square Kilometre Array Pathfinder (ASKAP) and demonstrated
a new calibration scheme for its phased array feeds, using an on-dish
calibration system and self-calibration.

{\bf Philipp Arras} explained Bayesian Radio Aperture Synthesis, which
reconstructs radio images of the sky using Bayesian inference.  This
is only possible using prior assumptions to reconstruct the sky \cite{arrasetal2018}. He is
further developing the software package RESOLVE, and shows that even
with a bad uv-coverage or considerable noise, the algorithm returns
valid sky maps of example source Cygnus~A.

\subsection{Cosmology, Large-scale Structure, and Galaxy clusters}

Much of the magnetism in the Universe is believed to be amplified
and/or maintained by some dynamo process. However, these dynamo
processes need tiny magnetic fields, called {\it seed fields}, to
amplify to observed field strengths and configurations. These seed
fields could have been created in the early universe by various
processes such as the Biermann battery, cosmological defects and
others as described below. Magnetic fields created in this era are
called {\it primordial magnetic fields}. Alternatively, it is possible
that the Universe got magnetized only later, e.g.\ at the time of the
reionization of the Universe.

Seed fields are of exceedingly small strength
($\sim 10^{-10} - 10^{-20}$~G), so will have to be amplified to the
cosmic magnetic fields we can observe today by a dynamo. Both the
process of the magnetization of the Universe and its occurrence in
time are still unknown, as are the exact dynamo processes in various
sources. A number of theories for creating and amplifying cosmic
magnetic fields were discussed at this conference, as well as the
possibility of observing primordial magnetic fields.

\subsubsection{Generation of cosmic magnetic fields}

{\bf Mathieu Langer} described a mechanism to create seed fields
during the Epoch of Reionization. In the first Str\"omgren spheres
created around the first generation of ionizing sources in the
Universe, energetic photons will escape the spheres and create a local
electric field. Inhomogeneities in the medium will induce a rotational
component in the electric field, which generates magnetic fields
\cite{durrivelanger2015, langeretalFT}. Ionizing sources can be the first (Pop III)
stars, ionizing galaxies or quasars, which would ionize on kpc to Mpc
scales. Therefore, this process can contribute to the large-scale
magnetization of the Universe.

Detection of intergalactic magnetic fields could prove an early
magnetization of the Universe. {\bf Teppei Minoda} discussed a
possible observational signature of primordial magnetic fields in
Cosmic Microwave Background (CMB) anisotropies, induced by the thermal
Sunyaev-Zel'dovich effect \cite{minodaetalFT}. He calculates the
evolution of gas temperature and density in the dark ages including
magnetic fields and finds a potentially detectable imprint of
primordial magnetic fields on the CMB at large multipoles $\ell \sim
10^5 - 10^6$.

{\bf Kerstin Kunze} presented numerical simulations of cosmological
magnetic fields and their influence on the reionization history of the
Universe \cite{kunze2018, kunzeFT}. She showed that high primordial
magnetic field values cause reionization to happen earlier. In the
particular examples studied, primordial magnetic fields would leave
observational traces in the 21cm line signal at frequencies above 120~MHz, which
could be observable with the Square Kilometre Array (SKA).

\subsubsection{Amplification of cosmic magnetic fields}

Amplification of cosmic seed fields to the values that we currently
observe requires a dynamo, converting various other sources of energy
into magnetic energy. The exact mechanisms, drivers, energy balance and
time scales are still under discussion.

{\bf Dongu Ryu} presented numerical simulations of turbulent dynamo
action in galaxy clusters, and showed that this gives a coherence
length of maximally $\sim 50$~kpc. Coherent magnetic fields in
clusters are observed to scales of $\sim 1$~Mpc, so that another
dynamo mechanism is needed to transport magnetic energy to larger
spatial scales. A possibility explored is {\it sporadically driven}
turbulent dynamo, which only flares up during galaxy mergers. This
will increase the injection length somewhat, but still cannot explain
the largest observed scales.

{\bf Jennifer Schober} discusses the amplification of magnetic fields
due to the Chiral Magnetic Effect. This effect can amplify magnetic
fields due to unequal numbers of left-handed and right-handed
fermions, which happens in special circumstances such as in
proto-neutron stars, in heavy-ion collisions in Earth-based colliders,
or in the first second of the Universe. Her numerical simulations show
initial exponential amplification of small-scale magnetic fields,
followed by a slower evolution of fields on significantly larger
spatial scales by turbulence developing \cite{schoberetal2018}.

{\bf Julius Donnert} argued that current large cosmological numerical
simulations need to be expanded. Magnetic fields amplified by the
turbulent dynamo grow from small scales to large scales, so one needs
to resolve the smallest scales ($\sim 3$~kpc), which is sub-grid for
many current cosmological models.
He introduced the Wombat software, which is a complete redesign of a
cosmological code which runs massively parallel on a supercomputer,
developed in collaboration with Cray Supercomputers \cite{donnertetalFT}.

\subsubsection{Magnetic fields in galaxy clusters}

Radio emission is ubiquitous in clusters, and can typically be divided
into two components: (1) radio halos in the centers of clusters, which
are roughly co-located with thermal X-ray emission and generally
unpolarized; and (2) radio relics at the outskirts of clusters, which
are highly polarized and believed to be created by large-scale shocks
due to galaxy mergers in the cluster.

{\bf Soonyoung Roh} investigates if numerical simulations can
reproduce observed radio halos and relics. The similarity between
radio and X-ray power spectrum slopes in halos suggests a direct
correlation between the thermal and non-thermal emission, indicating
re-acceleration of electrons. She probes various energy relations and
finds that a cosmic ray energy density comparable to the magnetic
energy density reproduces radio halos the best. Her simulations of
relics do reproduce shock regions with enhanced synchrotron emission.

{\bf Francesca Loi} simulates full-Stokes radio emission from galaxy
clusters including discrete foreground and background sources, to
investigate detectability of polarization in the clusters with the
SKA \cite{loietalFT}. She finds that the behavior of cluster radio
halos is a dominant factor significantly influences the Faraday depth
spectrum, and that Faraday screens and additional sources introduce
ambiguities in the Faraday spectrum which complicates interpretation.

{\bf Valentina Vacca} presents numerical simulations of synchrotron
emission from galaxy clusters, which allow estimating a lower limit to
the required magnetic field strength in clusters from non-detections
of halo polarization \cite{vaccaetalFT}. Her simulations confirm that with current
observational capabilities, most radio halos should be completely
depolarized. Decomposition methods allow distinguishing between the
Milky Way foreground and extragalactic background, but not the
distinction between the various extragalactic components (filaments,
galaxy clusters, voids, sheets) except in the very local Universe. New
1.4~GHz polarimetric observations with the Sardinia Radio Telescope
show a new population of radio halos that are larger than the known
halos and weaker in X-rays \cite{vaccaetal2018}.

{\bf Valentina Vacca} presents advanced techniques for the study of
magnetic fields in the large scale structure of the Universe, using
radio halos to study magnetic field strength and structure on scales
up to several hundreds of kpc. In addition, new 1.4~GHz polarimetric
observations with the Sardinia Radio Telescope show a new population
of diffuse synchrotron sources fainter and larger than known cluster
sources, potentially associated with magnetic fields in the cosmic web
\cite{vaccaetal2018}. Lastly, decomposition methods using Faraday
rotation of background radio galaxies allow statistical distinction
between the Faraday rotation due to the Milky Way, filaments, galaxy
clusters, voids, sheets, and consequently investigation of extragalactic
magnetic field properties \cite{vaccaetalFT}.

{\bf Hiroki Akamatsu} discussed the synergy between X-ray and radio
observations in galaxy clusters. He determined Mach numbers of shocks
in radio relics from the temperature profile across the shock, as
observed e.g.\ in the Sausage relic with the Suzaku
satellite. Assuming that diffuse shock acceleration is the dominant
electron acceleration process in these shocks, one can derive the Mach
number from radio observations too. In the Sausage, the two estimates
of the Mach number are in agreement. However, diffusive shock
acceleration cannot accelerate the electrons to the required energies,
which indicates that reacceleration of the electrons must play a role.

This is also shown by {\bf Motokazu Takizawa}, who concludes that the Mach
numbers derived from X-ray and radio emission in the Toothbrush relic
do {\it not} agree with each other. He also reports the detection of a
surface brightness edge in the cluster RXC~J1053.7+5453, which he
suggests may be due to a contact discontinuity \cite{itahanaetal2015, itahanaetal2017,takizawaetalFT}.

Detailed analysis on small-scale structure in the 'handle' of the
Toothbrush relic is presented by {\bf Matthias Hoeft}. The ridge of
the shock is frequency dependent and therefore cannot be due to a
magnetic enhancement but must instead be due to a traveling shock
front which cools. The narrowness of the peak combined with the
turbulence indicates that magnetic fields here are lower than
average. The fact that the Toothbrush relic is still highly polarized
at 3.6cm Effelsberg observations (with very low resolution) indicates
that the field across the relic is very regular.

\subsubsection{Intergalactic Magnetic Fields}

Intergalactic space is believed to be magnetized to some level as
well. In addition to theoretical estimates, the first tentative
detections of magnetic fields in intergalactic space and/or filaments
of the cosmic web are being discussed. Various ways to constrain or
predict intergalactic magnetic fields were discussed.

{\bf Takuya Akahori} studied the optimum frequency range to detect
intergalactic magnetic fields with Faraday tomography. He concluded
that for fairly small bandwidths around the optimum $500-600$~MHz,
intergalactic Faraday depths of $\sim 5$~rad~m$^{-2}$ should become
detectable \cite{akahorietal2018, akahorietalFT}.

{\bf Vikram Ravi} reviewed Fast Radio Bursts (FRBs) and discussed how
their rotation measures can help constrain progenitor models and probe
intergalactic space. Dispersion measures of FRBs measure the baryon
content of the intergalactic medium. Current observations can put an
upper limit of extragalactic magnetic fields of $\sim 20$~nG.

{\bf Justin Bray} uses arrival directions of Ultra-High Energy Cosmic
Rays (UHECRs) as probes to constrain intergalactic magnetic
fields. Assuming various plausible source populations for the UHECRs,
and neglecting deviations due to the Galactic magnetic field, he
arrives at a conservative upper limit for the intergalactic magnetic
field in cosmic voids of $\sim 0.1$~nG, which is lower than the current
CMB limits \cite{brayscaife2018}.

{\bf Shane O'Sullivan} re-imaged polarized point sources detected in
the HETDEX field (see Section~\ref{s:mw}) at high resolution with LOFAR. He finds
small Faraday depth differences between lobes in radio galaxies on Mpc
scales. This can be explained by an intergalactic magnetic field with
an rms strength $B_{rms} ~ 0.3~\mu$G. However, the poorly constrained
small-scale fluctuations in the Faraday depth of the Milky Way limit
the analysis \cite{osullivanetal2018, osullivanetalFT}.

\subsection{Galaxies and AGN}

\subsubsection{AGN}

{\bf Shane O'Sullivan} presents millimeter-spectropolarimetry results
of high Faraday depths ($\sim 10^5$~rad~m$^{-2}$) from the radio jet
of 3C~273 using Atacama Large Millimeter Array (ALMA) observations
(project led by Talvikki Hovatta \cite{hovattaetal2018}).

{\bf Craig Anderson} demonstrates that the thermal plasma in radio
galaxy lobes can be probed using Faraday Tomography \cite{andersonetalFT}. Observations of
Fornax~A at 1-3~GHz with the Australia Telescope Compact Array (ATCA)
show complex Faraday depth structure, with frequent small-scale
reversals of Faraday depth sign. Ruling out both Kelvin-Helmholtz and
Rayleigh-Taylor instabilities, he finds that the most likely
explanation for these structures is material advected from the host
galaxy NGC~1316. He also presents a polarization map of part of the
Southern lobe of Centaurus~A with the Australia SKA Pathfinder
(ASKAP), which shows complex depolarization structure.

The catalog of rotation measures (RM) of polarized NVSS sources
\cite{tayloretal2009} contains a few dozen sources at high latitudes
with anomalously high RMs. {\bf Yik Ki (Jackie) Ma} investigated the
cause of these anomalously high RMs using follow-up observations at
$1-2$~GHz with the VLA. Some of these RMs were caused by wrongly
evaluated $n\pi$-ambiguities. He suggests using the variation in RM in
a radius of $3^{\circ}$ as a diagnostic to find sources with a wrong
evaluation of the $n\pi$-ambiguity.

{\bf Alice Pasetto} performed spectro-polarimetric imaging of
14~polarized AGN with high RMs (RM~$>500$~rad~m$^{-2}$), using the
Jansky Very Large Array (JVLA) at 4~and 12~GHz. These sources show
widely varying total intensity synchrotron spectra and polarization
behavior. QU-fitting reveals multiple Faraday depth components with
Faraday depths from 100s to 1000s of rad~m$^{-2}$. These high-Faraday
depth components could be e.g.\ due to jet winds, or a clumpy medium
very close to the central black hole.

The role of magnetic fields in dust tori around Active Galactic Nuclei
(AGN) was discussed by {\bf Yuki Kudoh}, who showed that numerical
simulations of AGN tori need both high order accuracy and high
resolution in order to determine the role of the magneto-rotational
instability in the torus \cite{kudohwadaFT}.

{\bf Tomohisa Kawashima} investigated the influence of spin on the
shadow of the supermassive black hole in the center of M87, which will
be imaged by the Event Horizon Telescope (EHT). His general
relativistic MHD (GR-MHD) simulations show that very high spins may be
detectable by a slight increase in shadow radius.

\subsubsection{Spiral galaxies}

Radiopolarimetric observations of a large selection of edge-on spiral
galaxies, as part of the CHANG-ES project \cite{irwinetal2013}, were
shown by {\bf Marita Krause}.  She presented estimates of radio scale
heights for this sample, and concluded that gravitationally influenced
galactic winds are ubiquitous in spiral galaxies \cite{krauseetal2018}. She showed that the
scale heights of the polarized halos depend mainly on the diameter of
the radio disks of the galaxies. Lack of frequency dependence of scale
heights implies that cosmic ray transport is dominated by escape
through Galactic winds.

Numerical magnetohydrodynamical simulations of spiral galaxies by {\bf
  Mami Machida} show that weak fields amplified by the
magneto-rotational instability (MRI) become turbulent and eventually
form outflows driven by magnetic pressure \cite{machidaetalFT}. Therefore, a dynamo is
created produced by the MRI and the Parkes instability. She calculated
observable Faraday depths and polarized intensity from the simulation
results, including depolarization effects. The butterfly pattern as in
the RM map from the Northern VLA Sky Survey \cite{tayloretal2009} can
be reproduced, as well as magnetic arms and low-frequency
depolarization in the disk.

{\bf Hiroyuki Nakanishi} derived RMs for 6 bright galaxies at
inclination angle $i < 70^{\circ}$ from VLA data at 1.4~GHz and
4.8~GHz. He combined the magnetic field direction information from the
RMs with polarization vector orientations from synchrotron
polarization maps to obtain maps of integrated magnetic field vectors
across these galaxies. In all galaxies, small (or meso-)scale structure
in magnetic field can be seen, with abundant field reversals.

{\bf Kohei Kurahara} showed that the directions of magnetic field vectors
in nearby galaxy NGC~6946 as derived from radiopolarimetric maps show
reversals in field directions on meso-scales, which cannot be
explained by recent numerical simulations of galactic magnetic fields
developing from an axisymmetric magnetic field configuration \cite{kuraharaetalFT}.

{\bf Sarrvesh Sridhar} presented his work on the curious galaxy
NGC~4258, which shows anomalous arms in radio, X-rays and H$\alpha$
that are offset from the star forming disk. His Westerbork Synthesis
Radio Telescope (WSRT) continuum and HI observations and LOFAR data
show the anomalous arms. He finds a $\sim 6\%$ increase in polarized
intensity at a location of an HI hole in the western anomalous arm,
suggesting interaction between the polarizing gas and the star forming
disk. This leads him to propose a model where the western arm is
located in the plane of the star forming disk, but the eastern arm is
protruding at an angle form the disk.

{\bf Maja Kierdorf} presented S-band ($2-4$~GHz) JVLA polarimetric
data of nearby spiral M51,complementing existing data at
$\sim 1.5$~GHz, 5~GHz and 8~GHz. This intermediate frequency band is
ideal to probe the disk-halo interaction in M51. She showed that her
data cannot be fit well with published depolarization models for
M51 \cite{shneideretal2014},
so that a more complex interpretation is needed.

\subsection{Magnetic Fields in the Milky Way}
\label{s:mw}

\subsubsection{Magnetic fields in the general interstellar medium}

{\bf Cameron Van Eck} discussed LOFAR Faraday Tomography of a
$\sim 300$~square degree region at high latitudes, probing the nearby
interstellar medium (ISM) \cite{vanecketalFT}. He finds ubiquitous and
large-(angular-)scale polarized emission at various low
Faraday depths. His calculations of depolarization in
Faraday thick media show that Gaussian shapes get depolarized rapidly
at low frequencies. The turbulent ionized medium can be represented by
a Gaussian in Faraday space, leading him to the conclusion that LOFAR
is only sensitive to polarized emission building up in neutral
(Faraday-thin) clouds.

The Global Magneto-ionic Medium Survey (GMIMS, \cite{wollebenetal2009}) consists of six
individual surveys together covering two hemispheres and full
frequency coverage between $\sim 300$~MHz and $\sim 1800$~MHz. Two of
these surveys have been finished to date and their results were
discussed in this meeting.

Firstly, {\bf Alex Hill} presented the GMIMS-North-High Band study
($1300-1800$~MHz). He studied a highly polarized region in the outer
Galaxy called the Fan Region. Although it was long believed that the
Fan Region was local (a few 100 pc), depolarization of part of the Fan
region by the supernova remnant W4 now places (part of) the Fan region
firmly in and/or behind the Perseus Arm. Modeling (de-)polarization
due to large-scale magnetic field models confirms the presence of this
region \cite{hillFT}. He introduced the new Canadian Hydrogen Intensity Mapping
Experiment (CHIME) instrument, including a first preliminary image of
the whole northern sky.

The second GMIMS presentation focused on HII regions in the
GMIMS-South-Low survey ($300-480$~MHz). {\bf Alec Thomson} shows that
HII regions may completely depolarize background emission, or not
depolarize at all, depending on their distance and thermal electron
density. Depolarized HII regions can be used to estimate the
emissivity of their foregrounds.

In order to translate Faraday depth measurements to magnetic field
knowledge, a good model for thermal electron density in the Milky Way
is crucial. Dispersion Measures (DMs) of pulsars are an important
probe, which is often hampered by large uncertainties in their
distances. {\bf Osamu Kameya} discussed differential VLBI to determine
parallax distances. Using these, he finds that the thermal electron
density in the Solar neighborhood is quite complex and not accurately
represented in the NE2001 galactic electron density model.

Polarization of near infrared and optical starlight is an independent
tracer of the Galactic magnetic field. This method was discussed by
{\bf Tetsuya Zenko}, who observed 52 Cepheids towards the inner Galaxy
using the Infrared Survey Facility (IRSF) and added 14 Cepheids from
the literature. In a field towards the inner Galaxy, he can
reconstruct a global magnetic field parallel to the plane, but also
detects small-scale variations.

Understanding the interstellar medium includes understanding the
fragmentation of gas. Planary structures (filaments) exist at all
scales and epochs, oftentimes magnetic field induced. Their clumpy
fragmentation has been described numerically, but is analytically a
tough problem to solve. {\bf Jean-Baptiste Durrive} presented his
method to solve the fourth order gravitational instability, by
reducing it to an iterative second order problem. His solution agrees
very well with the numerical results.

\subsubsection{Magnetic fields in Galactic objects}

The Local Bubble is a superbubble blown by a collection of supernovae
a few million years ago, in which our Sun is embedded. By deriving an
analytical model for the magnetic field in the shell of the Local
Bubble, represented by an inclined spheroid off-centred from the Sun,
and comparing its predicted dust polarisation to that measured by
Planck at high Galactic latitudes, {\bf Marta Alves} finds that the magnetic
field in the local interstellar medium has been highly deformed by the
Local Bubble \cite{alvesetal2018}. 

Supernova remnants (SNRs) are believed to be the main Galactic sources
for acceleration of electrons to cosmic-ray energies. The most
plausible model used is Diffusive Shock Acceleration (DSA). However,
there are still many open questions, one of which is the unknown
cosmic ray acceleration efficiency. H$\alpha$ emission from SNR shock
fronts probe shock acceleration physics, argues {\bf Satoru
  Katsuda}. In particular, a difference in polarization degree in
narrow and broad H$\alpha$ line may be proportional to the
acceleration efficiency. He measured H$\alpha$ polarization in broad
and narrow H$\alpha$ lines in the young Tycho SNR, known to be a very
efficient particle accelerator. He finds equal polarization in broad
and narrow lines, consistent with interstellar polarization. {\bf Jiro
  Shimoda} studies correlation functions in 1.5~GHz VLA synchrotron
total intensity in the Tycho SNR, which is correlated to the magnetic
field power spectrum. The outer shell is consistent with a Kolmogorov
magnetic field power spectrum, but the inner edge of the shell is
not. He suggests that this is due to the contact discontinuity that is
located there. 

SS~433 is a (so far) unique Galactic object consisting of a stellar jet
protruding through the SNR W50. {\bf Haruka Sakemi} obtained ATCA
observations at $2.3-3.0$~GHz of SS~433 and showed that there is a
strong polarized filament behind the SS~433 jet head, and weaker
filaments behind it possibly indicating a helical magnetic
field \cite{sakemietalFT}. Faraday Tomography reveals a complex pattern of Faraday depth
components from $0$~to $\sim300$~rad~m$^{-2}$. This structure is
compared to the 2-temperature MHD simulations of astrophysical jets
presented by {\bf Takumi Ohmura}, which is the first jet simulation to
use separate electron and ion temperatures. She finds that the
electron temperature is 1-2 orders of magnitude lower than the ion
temperature at the jet hotspot, where the magnetic field is amplified
by the termination shock \cite{ohmuraetalFT}.

{\bf Mariko Nomura} explains the 'Bullet' in W44, a Y-shaped bright
emission feature in the SNR W44, by a bow shock from an isolated
stellar-mass black hole through the molecular cloud associated with
W44. Her MHD simulations show that only a magnetic field of 0.5~mG or
more can create a feature that agrees with the size and shape of the
observed feature. Three observed Compact High Velocity Clouds near the
Galactic Center may also have been created this way. Searching for
their interactions with molecular clouds may thus be a way to uncover
the hidden population of Galactic stellar-mass black holes.

{\bf Hiroyuki Takahashi} presented GR-MHD simulations of supercritical
accretion (i.e.\ accretion above the Eddington limit) onto a
magnetized neutron star by a binary companion star. He showed the
formation of a magnetosphere in these simulations and presented a
theoretical model of the magnetosphere radius and spin-up rate of the
central neutron star, consistent with the GR-MHD results.

{\bf Kenji Nakamura} presented 2D MHD simulations of the transitions
of X-ray binaries between accretion states. Simulations including
thermal conduction formed a low-temperature, high-density accretion
disk similar to simulations without thermal conduction, but in
addition produced a warm accretion flow around this accretion disk \cite{nakamuraetalFT}.

\subsubsection{Large-scale models of the Galactic magnetic field}

There are a fair number of global Galactic magnetic field models,
either 2D (only the disk) or 3D (including the gaseous halo). These
models are mostly fitted to synchrotron total intensity, polarized
intensity and/or RMs from pulsars and/or extragalactic sources. The
quality of fit of the latest models is somewhat comparable, and a
model sufficiently reliable to use for ISM studies or foreground
subtraction is not available yet. Improvements to these models were
discussed in the form of adding tracers and diagnostics, or improving
the modeling method.

{\bf Jennifer West} searches for non-zero helicity in the Milky Way,
which is predicted to cause skewness of the distribution in the RM vs
polarized intensity distribution. For this, she compares predictions
from dynamo models containing different
modes \cite{henriksenetal2018},
with synchrotron polarization data from the Planck satellite.

An other independent tracer of Galactic magnetic fields are OH masers,
found in many environments. However, when using them as probes of the
Galactic magnetic field, it is best to look in star forming regions,
where OH masers form at the outer edges.

{\bf Chikaedu Ogbodo} presented Zeeman splitting measurements of the
1720~MHz transition of hydroxyl masers as part of the MAGMO project \cite{greenetal2012},
and discussed polarimetric properties and derived magnetic field
strengths.

{\bf Jimi Green} discusses current and future polarimetric OH (hydroxyl) maser surveys to
detect magnetic fields in star forming regions through Zeeman
splitting. He also introduces the CH (carbine) maser as a tracer of a
gas phase between CO and atomic H, which enables measuring the
so-called 'dark magnetism'.

{\bf Marijke Haverkorn} presented the Interstellar MAGnetic field
INference Engine (IMAGINE), which is a software package to model the
Galactic magnetic field using Bayesian inference, as well as the
associated (open) collaboration of researchers in various fields
interested in Galactic magnetism \cite{haverkornetalFT,boulangeretal2018}.

\subsection{Amazing Magnetism Projects}

In this session, a few of the exciting instruments or surveys expected
to enable breakthroughs in cosmic magnetism in the near future were
discussed.

{\bf Bryan Gaensler} presented the first results from Wide Field
Polarization Surveys. Firstly, the ASKAP POlarisation Sky Survey of
the Universe's Magnetism (POSSUM) survey is an all-southern sky
polarimetric survey at $1.1 - 1.4$~GHz at 10~arcsecond resolution and
down to $\sim20~\mu$Jy/beam sensitivity. Secondly, the Very Large
Array Sky Survey (VLASS) will survey the northern sky from 2 to 4~GHz
at 2.5~arcsec resolution to a sensitivity of $\sim70~\mu$Jy/beam. A
major goal of these surveys is the creation of a RM Grid of
extragalactic point sources, which allows study of magnetic fields in
the Milky Way, external galaxies, AGN, clusters, and the intergalactic
medium.

{\bf Tessa Vernstrom} gave an overview of the Murchison Widefield
Array (MWA), and a wide variety of MWA science results including
limits on synchrotron radiation from the cosmic web
\cite{vernstrometal2017} or absorption studies of Galactic HII regions
\cite{hindsonetal2016}. She presented the GaLactic and Extragalactic
All-Sky MWA Survey (GLEAM, \cite{waythetal2015} catalog,
visualization, science results, and follow-up surveys.

{\bf George Heald} presented both the LOFAR Multifrequency Snapshot
Sky Survey (MSSS) and the Galactic and Extragalactic All-sky MWA
(GLEAM) survey, that together form a truly all-sky polarimetric source
catalog at low frequencies. He showed early MSSS results and discussed
the survey's upgrade to 45~arcsec resolution. The polarization
analysis of the MSSS data, the MSSS All-sky Polarization Survey
(MAPS), was presented by {\bf Jamie Farnes}.  The GLEAM results
include ongoing analysis of polarized point sources, the GLEAM source
catalog and the MWA's upgrade to longer baselines.

{\bf Jamie Farnes} also presented the the SKA Science Data Processor
Integration Prototype (SIP), which is an end-to-end prototype of the
major components of the Science Data Processor of the SKA. He
discussed inclusion of the MAPS data pipeline into SIP including
Faraday Tomography, and showed early results of MAPS data as processed
with SIP \cite{farnesetalFT}.



\section{Conclusions}

This highly successful workshop saw presentations of broad and varied
scientific results on many aspects of cosmic magnetism, from
observations, simulations, as well as theory. New results on cosmic
magnetism from stellar scales to large-scale structure formation, and
from the magnetic field generation in the early Universe to magnetism
in nearby Galactic objects were presented and discussed. 

As the Japanese part of the Scientific Organizing Committee (SOC) was
a representation of the Japan SKA Consortium (SKA-JP) Cosmic Magnetism
Science Working Group, an important topic in the workshop was the
future of radio astronomy and radio-polarimetry in particular, as
addressed in many presentations.  The audience was presented with
exciting new observations of various new or upgraded radio telescopes
and instruments, such as the CHIME, LOFAR, SRT, MWA, and ASKAP
telescopes and VERA, ATCA-CAB and Parkes PAF receivers. In addition to
observational advances, developments in computing are equally
important. Updates were presented on numerical methods to process and
(statistically) analyze complex data sets, computing hardware and
software developments for the SKA, and to prepare for the era of
exascale computing.

\vspace{6pt}


\authorcontributions{writing—original draft preparation, M.H.;
  writing—review and editing, all; SOC chairing, M. M. and M. H.;
  oral conference summary, T. A.}
%

\funding{A part of this conference was achieved using the grant of
  Research Assembly supported by the Research Coordination Committee,
  National Astronomical Observatory of Japan (NAOJ), National
  Institutes of Natural Sciences (NINS). This conference was supported
  by Miyazaki Convention \& Visitors Bureau. The project leading to
  this publication has received funding from the European Union’s
  Horizon 2020 research and innovation programme under grant agreement
  No 730562 [RadioNet]. MH acknowledges funding from the European
  Research Council (ERC) under the European Union’s Horizon 2020
  research and innovation programme (grant agreement No 772663).}

\acknowledgments{The authors would like to thank all participants to the
workshop and all contributors to this Special Issue for their
contributions to make this workshop such a successful event.}

\externalbibliography{yes}
\bibliography{references}



\end{document}